\documentclass[make,article,accept,moreauthors,pdftex]{Definitions/mdpi}

\usepackage{multirow} 
\usepackage{graphicx} 
\usepackage[labelformat=simple]{subfig}

\usepackage{booktabs} 
\usepackage{listings}
\usepackage{algorithm}
\usepackage{algorithmic}
\usepackage{multicol}
 
\newlength\hrulethickness
\makeatletter
\renewcommand\fs@ruled{\def\@fs@cfont{\bfseries}\let\@fs@capt\floatc@ruled
  \def\@fs@pre{\hrule height \hrulethickness depth0pt \kern2pt}%
  \def\@fs@post{\kern2pt\hrule height \hrulethickness depth0pt \relax}%
  \def\@fs@mid{\kern2pt\hrule height \arrayrulewidth depth0pt \kern2pt}%
  \let\@fs@iftopcapt\iftrue}
\makeatother

\usepackage[normalem]{ulem}
 \usepackage{ulem}

\usepackage{xcolor} 
\definecolor{blue}{RGB}{51,153,255}

\usepackage{amsmath}

\definecolor{eclipseStrings}{RGB}{42,0.0,255}
\definecolor{eclipseKeywords}{RGB}{127,0,85}
\colorlet{numb}{magenta!60!black}

\lstdefinelanguage{json}{
basicstyle=\normalfont\ttfamily,
commentstyle=\color{eclipseStrings}, 
stringstyle=\color{eclipseKeywords}, 
numbers=left,
numberstyle=\scriptsize,
stepnumber=1,
numbersep=8pt,
showstringspaces=false,
breaklines=true,
frame=lines,
backgroundcolor=\color{white}, 
string=[s]{"}{"},
comment=[l]{:\ "},
morecomment=[l]{:"},
literate=
*{0}{{{\color{numb}0}}}{1}
{1}{{{\color{numb}1}}}{1}
{2}{{{\color{numb}2}}}{1}
{3}{{{\color{numb}3}}}{1}
{4}{{{\color{numb}4}}}{1}
{5}{{{\color{numb}5}}}{1}
{6}{{{\color{numb}6}}}{1}
{7}{{{\color{numb}7}}}{1}
{8}{{{\color{numb}8}}}{1}
{9}{{{\color{numb}9}}}{1}
}

\continuouspages{yes}
\firstpage{333} 
\pubvolume{3}
\issuenum{2}
\articlenumber{17}
\pubyear{2021}
\copyrightyear{2021}
\externaleditor{{Academic Editor}: Edgar Weippl}
\datereceived{30 January 2021}
\dateaccepted{24 March 2021 }
\datepublished{29 March 2021}
\hreflink{https://doi.org/\linebreak 10.3390/make3020017} 

\Title{Privacy and Trust Redefined in Federated Machine Learning}
\TitleCitation{Privacy and Trust Redefined in Federated Machine Learning}

\Author{Pavlos Papadopoulos 
$^{1,}$*\href{https://orcid.org/0000-0001-5927-6026}{\orcidicon}, Will Abramson $^{1}$, Adam J. Hall $^{1}$, Nikolaos Pitropakis $^{1,2}$\href{https://orcid.org/0000-0002-3392-9970}{\orcidicon} and William J. Buchanan $^{1}$\href{https://orcid.org/0000-0003-0809-3523}{\orcidicon}}

\AuthorNames{Pavlos Papadopoulos, Will Abramson, Adam J. Hall, Nikolaos Pitropakis and William J. Buchanan}
\AuthorCitation{Papadopoulos, P.; Abramson, W.; Hall, A.J.; Pitropakis, N.; Buchanan, W.J.}

\address{%
$^{1}$ \quad Blockpass ID Lab, School of Computing, Edinburgh Napier University, Edinburgh EH10 5DT, UK; will.abramson@napier.ac.uk (W.A.); {adam.hall@napier.ac.uk (A.J.H.)};
N.Pitropakis@napier.ac.uk or {nikolaos.pitropakis@8bellsresearch.com} 
(N.P.); B.Buchanan@napier.ac.uk (W.J.B.)\\
$^{2}$ \quad Eight Bells LTD, Nicosia 2002, Cyprus}
\corres{Correspondence: pavlos.papadopoulos@napier.ac.uk }

\abstract{{
A common privacy issue in traditional machine learning is that data needs to be disclosed for the training procedures. In situations with highly sensitive data such as healthcare records, accessing this information is challenging and often prohibited. Luckily, privacy-preserving technologies have been developed to overcome this hurdle by distributing the computation of the training and ensuring the data privacy to their owners. The distribution of the computation to multiple participating entities introduces new privacy complications and risks. In this paper, we present a privacy-preserving decentralised workflow that facilitates trusted federated learning among participants. Our proof-of-concept defines a trust framework instantiated using decentralised identity technologies being developed under Hyperledger projects Aries/Indy/Ursa. Only entities in possession of Verifiable Credentials issued from the appropriate authorities are able to establish secure, authenticated communication channels authorised to participate in a federated learning workflow related to mental health data.}}


\begin{document}

\section{Introduction}
{Machine Learning (ML) and Deep Neural Networks (DNN) gained popularity in the last few years due to technology advancement. ML and DNN infrastructures can analyse a vast amount of information to predict a certain case~\cite{canziani2016analysis,liu2017survey,holzinger2019causability}.} DNN is a part of ML that tries to replicate a human's brain neurons' functionalities to achieve a prediction. This analysis and prediction functionalities can deal with complex problems that were previously considered to be unsolvable. ML predictions become more valuable when the analysis involves highly sensitive private data such as health records. Consequently, data holders cannot simply share their private data with ML algorithms and experts~\cite{chen2012data}. Many defensive techniques proposed in the past as countermeasures to the information leakage of sensitive data, such as anonymisation and obfuscation techniques~\cite{zhang2018privacy}. {Other research focused on non-iterative Artificial Neural Network (ANN) approaches for data security~\cite{tkachenko2019model,IZONIN2020}.} Nevertheless, due to the advancement of technology, similar techniques cannot anymore guarantee the privacy of the underlying data~\cite{hall2016predicting,ahmad2019barriers}. Malicious users are able to reverse and reconstruct, anonymised and obfuscated data, in~order to identify the identities of the data~subjects.

It is common knowledge that data is the most valuable asset of our century. Since ML algorithms require vast amounts of it, it is frequently targeted by malicious parties. {Several attacks exist that can hack, reconstruct, reverse or poison ML algorithms~\cite{kairouz2019advances,FeatureCloud}.} The common goal of these attacks is to identify the underlying data. Several of them require access during the ML algorithm training to succeed, while others are able to interfere later in the testing or publication phase. In~the literature, there are several defensive methods and techniques proposed against the aforementioned mistreats. However, when reinforcing a ML algorithm with security and privacy features against adversarial attacks, there is an impact on efficiency, thus ending up with the produced predictions not related to the associated tasks and with often a lower accuracy. Hence, a~balance between tolerable defence and usability is a critical point of interest that many researchers are trying to solve. Most of the aforementioned attacks are applicable and target centralised ML infrastructures, in~which the model owners have access and acquire all the training data~\cite{munoz2017towards}.

Due to the importance of the ML field and the associated privacy risks, a~new division of ML was created, namely Privacy-Preserving ML (PPML) \cite{al2019privacy}. This area is focused on the advancement and development of countermeasures against information leakage in ML, by~shedding light on the privacy of the underlying data subjects. When a few of these ML risks and attacks were introduced, they were purely theoretical; hence, due to the rapid advancement and evolution of ML, attackers led to the exploitation of those weaknesses in order to breach, steal, and~profit from this data. Several techniques and countermeasures were proposed in the PPML field. It is a common belief that if data never leave their holders possession to be used from a ML algorithm, then the data privacy is higher. The~most extended and researched area related to this is Federated Learning (FL)~ \cite{mcmahan2016communication,konevcny2016federated,bonawitz2019towards,Ryffel2018}.

In a FL scenario, the~ML model owners are able to send their model to the data holders for training. From~a high-level perspective, this scenario is secure; however, there are still many security flaws that need to be solved~\cite{kairouz2019advances}. For~example, the~ML model owners could reverse their model and identify the underlying training data~\cite{fredrikson2014privacy,fredrikson2015model}. A~suggested solution is to use a secure aggregator, often an automated procedure or program, as~a middle-ware, which aggregates all the participants' trained models and then sends the updates to the ML model owners~\cite{DBLP:journals/corr/BonawitzIKMMPRS16}. This solution is robust against several attacks~\cite{fredrikson2014privacy,fredrikson2015model,song2017machine,shokri2017membership,salem2018ml}, but~still involves issues, such as the possibility of a Man-In-The-Middle (MITM) attack that is able to interfere and trick both parties or even the scenario where one or more participants are malicious. In~the latter scenario, malicious data providers can poison the ML model~\cite{bagdasaryan2018backdoor,bhagoji2018analyzing,liu2017trojaning} in order to miss-classify specific predictions in its final testing phase; thus, the~ML model owner could never distinguish a poisoned model from a benign. This model poisoning scenario could happen in a healthcare auditing scheme, where the trusted auditing organisation uses a ML algorithm to audit other healthcare institutions to predict financial profits and losses in the future. A~potential malicious healthcare institution is able to poison the ML model using indistinguishable data that could lead to false predictions from the final trained model by miss-classifying the economic losses of the healthcare institution and approve its operation, as~usual.

The aforementioned attacks share some common issues and concerns such as the lack of trust between the participating parties, or~the lack of a secure communication channel to transmit private ML model updates. In~this work, we redefine the privacy and trust in federated machine learning by creating a Trusted Federated Learning (TFL) framework as an extension of the privacy-preserving technique to facilitate trust amongst federated machine learning participants~\cite{abramson2020distributed}. In~our scheme, the~ML participants need to get a certification before their participation, from~a trusted governmental body such as the National Health Service (NHS) Trust in the United Kingdom. Following, the~ML training procedure is distributed among the participants similarly to FL. The~main difference is that the model updates are being sent through a secure communication end-to-end encrypted channel. Before~learning commences, the~respective parties must authenticate themselves against some predetermined policy. Policies can be flexibly determined based on the ecosystem, trusted entities and associated risk; this paper gives an example of using a healthcare scenario. {The proof-of-concept developed in this paper is built using open-source Hyperledger technologies such as Aries/Indy/Ursa~\cite{aries,indy,ursa} and developed within the PyDentity-\textit{Aries FL} 
project~\cite{PyDentity,Ariesfl} of the OpenMined open-source privacy-preserving machine learning organisation.} Our scheme is based on advanced privacy-enhancing attribute-based credential cryptography~\cite{camenisch2002signature,camenisch2013concepts} and is aligned with emerging decentralised identity standards; Decentralized Identifiers (DIDs) \cite{dids}, Verifiable Credentials (VCs) \cite{verifiable_creds} and DID Communication~\cite{didcomm}. The~implementation enables participating entities mutually authenticate digitally signed attestations (Credentials), issued by trusted entities specific to the use-case. The~presented authentication mechanisms could be applied to any regulatory workflow, data collection, and~data processing and are not limited solely to the healthcare domain. The~contributions of our work could be summarised as follows:
\begin{itemize}
\item We enable stakeholders in the learning process to define and enforce a trust model for their domain through the application of decentralised identity standards. We also extended the credentialing and authentication system by separating Hyperledger Aries agents and controllers into isolated entities.
\item We present a decentralised peer-to-peer infrastructure, namely TFL, which uses DIDs and VCs in order to perform mutual authentication and federated machine learning specific to a healthcare trust infrastructure. Development and evaluation of explicitly designed libraries for federated machine learning through secure Hyperledger Aries communication channels.
\item {We demonstrate performance improvement upon our previous trusted federated learning state-of-the-art without sacrificing the privacy guarantees of the authentication techniques and privacy-preserving workflows.}
\end{itemize}

Section~\ref{litreview} provides the background knowledge and describes the related literature. Furthermore, Section~\ref{methodology} outlines our implementation overview and architecture, followed by Section~\ref{evaluation}, in~which we provide an extensive security and performance evaluation of our system. {Finally, our work concludes with Section~\ref{conclusion} that draws the conclusions, limitations, and~outlines approaches for future work.}

\section{Background Knowledge and Related~Work}
\label{litreview}

Recent ML advancements can accurately predict specific circumstances using relevant data. Hence, that led businesses and organisations to collect vast amounts of data to predict a situation before their competitors. The~rationale is often to analyse people's behaviour patterns to predict the next trend they will follow~\cite{yeh2018pursuing}. However, the~European Union tried to minimise and constrain this massive collection of data with the General Data Protection Regulation (GDPR) legislation~\cite{voigt2017eu}.

Another field that can take advantage of the recent ML progression is the healthcare sector. However, in~that case, the~underlying data used for ML training is sensitive and private. Thus, its privacy must be ensured first prior to the improvement of its ML predictions. The~aforementioned procedure's complexity raises when it is being outsourced to a third-party organisation specialising on the ML task; since the ML practitioners have the expertise to solve the task, but~a healthcare organisation is holding the required sensitive~data.

\subsection{Trust and the Data~Industry}
\label{trustindataindustry}

The notion of trust has been defined as domain and context-specific since it specifies the amount of control a party provides to another~\cite{field_guide_trust,hoffman2002conceptualization}. It is often represented as a calculation of risk since it can only be restrained and not fully eradicated~\cite{keymolen2016trust}. Accordingly, patients trust healthcare institutions when giving their consent to collect their data. However, huge volumes of medical data can be valuable in-context to ML algorithms that aim to predict particular cures or~conditions.

In 2015, Royal Free London NHS Trust outsourced patients sensitive data to a third-party ML company, particularly DeepMind, to~train ML algorithms for the early detection of kidney failure~\cite{powles2017google,streams_google}. However, this sensitive data usage was not regulated, raised concerns about data privacy and later judged as illegal by the Information Commissioner's Office~\cite{streams_ico}. This misbehaviour did not cause other researchers to use sensitive data for ML predictions and led them to obtain proper authorisation from the Health Research Authority first and then use the sensitive medical records to analyse retinal imaging automatically~\cite{de2016automated}, and~the segmentation of tumour volumes and organs of risk~\cite{chu2016applying}.

\subsection{Decentralised~Identifiers}
\label{sectiondids}

Recently, Decentralised Identifiers (DIDs) were established as a digital identifier in a World Wide Web Consortium (W3C) working group~\cite{dids}, that can magnify trust in distributed environments. DIDs can be controlled solely by their owners and grant a person the ability to be authenticated similar to a login system, but~without relying on a trusted third-party company. Consequently, DIDs are often stored in distributed ledgers such as blockchain ledgers, which are not managed by a single authority. Distributed storage systems such as Ethereum, Bitcoin and Sovrin ledgers, or~InterPlanetary File System (IPFS) are often used to store DID specifications, each with their own resolution method~\cite{did_methods}. An~outline of a DID document that would have been resolved from \textit{did:example:123456789abcdefghi}, using the DID method \textit{example} and the identifier \textit{123456789abcdefghi}, can be seen in \mbox{Listing~1}
. A~DID document consists of:
\begin{itemize}
\item ID-the DID that resolves to this document
\item Public key
\item Authentication protocols
\item Service endpoints
\end{itemize}

\noindent  \hspace{12pt}
\begin{minipage}[b]{0.958\linewidth}
\captionsetup[lstlisting]{position=top,labelsep=period,singlelinecheck=false, margin=0pt, labelfont={bf, small, stretch=1.17},textfont={small, stretch=1.17}}
\label{didlist}

\begin{lstlisting}[language=json,caption={An example DID document.}, linewidth=\columnwidth, basicstyle=\small, breaklines=true ]
{
  "@context": "https://example.org/example-method/v1",
  "id": "did:example:123456789abcdefghi",
  "publicKey": [{
    "id": "did:example:123456789abcdefghi#keys-1",
    "type": "RsaVerificationKey2018",
    "controller": "did:example:123456789abcdefghi",
    "publicKeyPem": "-----BEGIN PUBLIC KEY...END PUBLIC KEY-----\r\n"
  }],
  "authentication": [
    "did:example:123456789abcdefghi#keys-1",
 ],
  "service": [{
    "id": "did:example:123456789abcdefghi#agent",
    "type": "AgentService",
    "serviceEndpoint": "https://agent.example.com/8377464"
  }]
}
\end{lstlisting}
\end{minipage}

DID specifications assure the interoperability across the DID schemes in order to interact and resolve a DID from any storage system. Nonetheless, Peer DIDs implementations are used in peer-to-peer connections that do not require any storage system, in~which each peer stores and maintains their own list of DID documents~\cite{peer_did}.

\subsubsection*{Decentralised Identifiers Communication~Protocol}
\label{didcomm}

Hyperledger Aries is an open-source project~\cite{aries}, that uses decentralised identifiers to provide a public key infrastructure for a set of privacy-enhancing attribute-based credential protocols~\cite{camenisch2013concepts}. Hyperledger Aries implements DID Communication (DIDComm) \cite{didcomm}, a~communication protocol similar to one first outlined by David Chaum~\cite{chaum1981untraceable}. DIDComm is an asynchronous encrypted communication protocol that uses information from the DID document, such as the public key and their associated endpoint, in~order to exchange secure messages; the authenticity and integrity of the messages are verifiable. DIDComm protocol is actively developed by the Decentralised Identity Foundation~\cite{didcomm_dif}.

An example using the DIDComm protocol can be seen in Algorithm~\ref{Algorithm:DID}, in~which Alice and Bob want to communicate securely and privately. Alice encrypts and signs a message for Bob. Alice's endpoint sends the signature and the encrypted message to Bob's endpoint. Bob can verify the message's integrity by resolving the DID and checking if it corresponds to Alice's public key, decrypt and read the message. All the associated information required for this interaction are defined in each person's respective DID document. The~encryption techniques used by DIDComm include ElGamal~\cite{elgamal1985public}, RSA~\cite{rivest1978method} and elliptic curve-based~\cite{wohlwend2016elliptic}.

\begin{algorithm}[h!]
\caption{DID Communication Between Alice and Bob~\cite{abramson2020distributed}}
\label{Algorithm:DID}
\begin{spacing}{1.2}
\begin{algorithmic}[1]
\STATE Alice has a private key $sk_a$ and a DID Document for Bob containing an endpoint ($endpoint_{bob}$) and a public key ($pk_b$).
\STATE Bob has a private key ($sk_b$), and~a DID Document for Alice containing her public key ($pk_a$).
\STATE Alice encrypts plaintext message ($m$) using $pk_b$ and creates an encrypted message ($e_b$).
\STATE Alice signs $e_b$ using her private key ($sk_a$) and creates a signature ($\sigma$).
\STATE Alice sends $(e_b, \sigma)$ to $endpoint_{bob}$.
\STATE Bob receives the message from Alice at $endpoint_{bob}$.
\STATE Bob verifies $\sigma$ using Alice's public key $pk_a$
\IF{Verify$(\sigma, e_b, pk_a) = 1$}
\STATE Bob decrypts $e_b$ using $sk_b$.
\STATE Bob reads the plaintext message ($m$) sent by Alice
\ENDIF
\end{algorithmic}
\end{spacing}
\end{algorithm}
\vspace{-6pt}

\subsection{Verifiable~Credentials}
\label{sectionvcs}

\begin{figure}[H]
\includegraphics[width=0.95\linewidth]{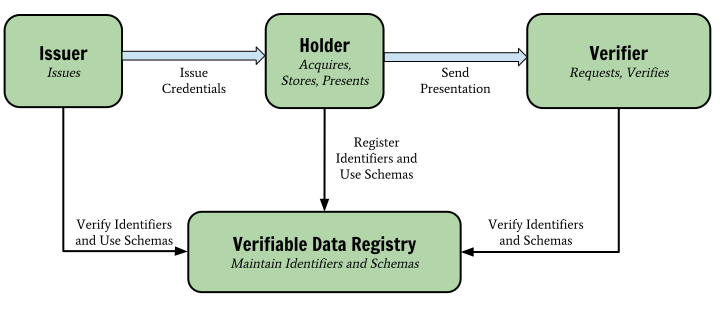}
\caption{Verifiable Credential Roles~\cite{verifiable_creds}.}
\label{fig:credential_roles}
\end{figure}

Verifiable Credentials (VCs) \cite{verifiable_creds}, is a set of tamper-proof claims that used by three different entities, \textit{Issuers}, \textit{Holders} and \textit{Verifiers}, as~it can be seen in Figure~\ref{fig:credential_roles}. VC model specification became a W3C standard in November 2019. A~distributed ledger is often used for the storage of the credential schemes, DIDs, and~Issuers' DID~documents.

The Issuer to create a new credential needs to generate a signature using their private key corresponded to their public key defined in their DID document. There are three valid categories of signature schemes such as  Camenisch-Lysyanskaya (CL) signatures~\cite{camenisch2002signature,cl_signature_2003}, Linked Data signatures~\cite{ld_sigs} and JSON Web signatures~\cite{json_web_sigs}. Hyperledger Aries uses CL signatures to create a blinded link secret, in~which credentials are tied to their intended entities by including a private number within them, without~the Issuers be aware of their values. It is a production implementation of a cryptographic system for achieving security without authentication first outlined in 1985~\cite{chaum1985security}.

The Verifier in order to accept the received credential from its Holder needs to confirm the following:
\begin{enumerate}
\item The Issuer's DID can be resolved to a DID document stored on the public ledger. The~DID document contains the public key that can be used to ensure the credential's~integrity.
\item The credential Holder can prove the blinded linked secret by creating a zero-knowledge proof to demonstrate it.
\item The issuing DID has the authority to issue this kind of credential. The~signature solely proves integrity, but~if the Verifier accepts credentials from any Issuers, it would be prone to obtain fraudulent credentials. It is possible to form a legal document outlining the operating parameters of the ecosystem~\cite{rfcToIP}.
\item The Issuer has not revoked the presented credential. This is done by checking that a revocation identifier for the credential is not present within a revocation registry (a cryptographic accumulator~\cite{fischlin_dynamic_2009}) stored on the public ledger.
\item Finally, the~Verifier needs to check that the credential attributes meet authorisation criteria in the system. It is common for a credential attribute to be valid only for a certain period.
\end{enumerate}{}

All the communication between the participating entities transmits peer-to-peer through a DIDComm protocol. It should be noted that a Verifier does not require to contact the credential's Issuer to verify a~credential.

\subsection{Docker~Containers}
\label{dockers}

All the participating entities presented in our work, take the form of Docker containers~\cite{boettiger2015introduction}. Docker containers are lightweight, autonomous, virtualised systems similar to virtual machines~\cite{smith2005architecture}. The~main difference between virtual machines is that Docker containers use the host's underlying operating system and bridge the network traffic in a virtual network card instead of being fully isolated. Moreover, Docker containers are being developed into deployable images that are executed and operate as expected invariably in any system that supports the Docker environment. Hence, applications that could be built using Docker containers are favourable for reproducibility and code replication purposes. However, since Docker containers use a virtual network card in their host machine's to redirect the network traffic, a~security testing in their ecosystem varies~\cite{martin2018docker}.

\subsection{Federated Machine~Learning}
\label{sectionfl}

FL can be expressed as the decentralisation of the ML. Opposed to centralised ML, in~a FL scenario, the~training data remain at their respective owners instead of transmitting to a central location to be used by a ML practitioner. There are several FL variations such as Vanilla FL, Trusted Model Aggregator, and~Secure Multi-Party Aggregation~\cite{kairouz2019advances,abramson2020distributed,DBLP:journals/corr/BonawitzIKMMPRS16,kholod2021open}. Consequently, the~ML model is primarily distributed among the data holders, who train it using their private data, and~then send it back to the ML model owner. ML training decentralisation permits data holders with sensitive data such as healthcare institutions to train useful ML algorithms to predict a cure or a disease. One of the FL advancements, namely Secure Multi-Party Aggregation, developed to further enhance the system's security by encrypting the models into multiple shares and aggregating all the trained models to eliminate the possibility of a malicious ML model owner~\cite{das2016distributed,Ryffel2018}.

To measure the accuracy of FL algorithm is similar to the traditional ML. Four metrics described in the list below~\cite{tharwat2020classification,shah2018performance}, are used for the calculation as follows:
\begin{enumerate}
\item True positive (TP): the model correctly predicts the positive prediction; correct
\item True negative (TN): the model correctly predicts the negative prediction; correct
\item False positive (FP): the model incorrectly predicts the positive prediction; false
\item False negative (FN): the model incorrectly predicts the negative prediction; false
\end{enumerate}
\newpage

The accuracy of the model is calculated from the number of \textit{TP}s and \textit{TN}s, 
divided by the total number of outcomes, as~you can see in Equation~(\ref{eqn:accuracy}).
\begin{equation}
\label{eqn:accuracy}
Model's Accuracy = \frac{(Correct Predictions: TP+TN)} {(Outcomes: TP+TN+FP+FN)}
\end{equation}

\subsection{Attacks on Federated~Learning}
\label{FLattacks}

Since that in FL, data never leaves their owners' premises, one could naively assume that FL is entirely protected against misuses. However, even if FL is more secure than traditional ML approaches, it is still susceptible to several privacy attacks that aim to identify the underlying training data or trigger a miss-classification on the final trained model~\cite{kairouz2019advances,buchanan2020review}.

Model Inversion attack~\cite{fredrikson2015model,fredrikson2014privacy,zhang2019secret}, is the first of its kind that aim to reconstruct the training data. A~potential attacker with access to the target labels can query the final trained model and exploit the returned classification scores to reconstruct the rest of the~data.

In Membership Inference attacks~\cite{shokri2017membership,salem2018ml}, the~attacker tries to identify if some data was part of the training. As with model inversion attacks, the~attacker exploits the returned classification scores in order to create several \textit{shadow} models that have similar classification boundaries as the original model under~attack.

In Model Encoding attacks~\cite{song2017machine}, the~attacker with white-box access to the model tries to identify the training data that have been \textit{memorised} by the model's weights. In~a black-box situation, the~attacker overfits the original training model in order for it to leak part of the target~labels.

From the other side, Model Stealing attacks~\cite{tramer2016stealing}, present the scenario of a malicious participant that tries to \textit{steal} the model. Since the model is being sent to the participants for training, malicious participants can construct a second model that mimics the original model's decision boundaries. In~that scenario, the~malicious participants could avoid paying usage fees to the original model's ML experts or sell the model to third~parties.

{Likewise, in~Model Poisoning attacks~\cite{bagdasaryan2018backdoor,bhagoji2018analyzing,liu2017trojaning,nuding2020poisoning}, since the malicious participants contribute to the training of the model, they are able to inject backdoor triggers to the trained model.} According to~\cite{bagdasaryan2018backdoor}, a~negligible number of malicious participants is able to poison a large model. Hence, the~final trained model would seem legitimate to the ML experts and react maliciously only on the given backdoor trigger inputs. In~that case, the~malicious participant could potentially \textit{trick} the original model when certain inputs are given. {Contrary to the Data Poisoning attacks~\cite{sun2020data,munoz2017towards,jagielski2018manipulating,biggio2012poisoning,laishram2016curie}, in~which the poison backdoor triggers are part of the training data, and~the ML model's accuracy may drop~\cite{steinhardt2017certified}.}

In Adversarial Examples~\cite{goodfellow2014explaining,carlini2017towards,chen2019hopskipjumpattack,papernot2017practical,yuan2019adversarial,pitropakis2019taxonomy}, the~attacker tries to \textit{trick} the model in order to classify falsely a prediction. The~threat model for this type of attacks is both white-box and black-box; thus the attacker does not require access to the training procedure, with~a potential attacking scenario to be a malware that evades the detection of a ML intrusion detection~system.

\subsection{Defensive Methods and~Techniques}
\label{defensive}

Due to the sensitivity of the underlying training data in ML, there are several defensive techniques, albeit several of them are still in the theoretical stage and therefore are not applicable~\cite{kairouz2019advances,abramson2020distributed}.

\subsubsection{Differential~Privacy}
\label{dp}

The most eminent defensive countermeasure against many privacy attacks in ML is Differential Privacy (DP), a~mathematical guarantee that ensures the ML algorithm's output, despite if a particular person's data used for the training procedure~\cite{dwork2008differential,dwork2011differential}. The~formal mathematical proof can be seen in Equation~(\ref{eqDP}), where the probability \textit{A} 
for all \textit{C} that are in range (\textit{A}), is differentially private, if~for any two adjusted databases \textit{D} and \textit{D'} that alter in only one element exists:
\begin{equation}
\label{eqDP}
P(A(D) \in C) \leq e^\varepsilon P(A(D') \in C)
\end{equation}

DP elaborates noise techniques to protect a ML algorithm from attacks; however, its accuracy drops significantly according to the designated privacy level~\cite{abadi2016deep}. Researchers, further extended and relaxed this mathematical proof into ($\epsilon$, $\delta$)-DP, which introduced an extra $\delta$ feature that limits the probability for errors~\cite{abadi2016deep,mcmahan2018general,dwork2014algorithmic,mironov2017renyi}.

\subsubsection{Secure Multi-Party~Computation}
\label{smpc}

Secure Multi-Party Computation (SMPC) \cite{goldreich1998secure}, is a cryptographic function that allows several participants to compute a procedure mutually, such a ML training procedure. Only the outcome of the function is disclosed to the participating parties and not the underlying training information. Using SMPC, gradients and parameters can be computed and updated encrypted in a decentralised manner. In~this case, each data item's custody is split into shares to be held by relevant participating entities. SMPC is able to protect ML algorithms against privacy attacks that target the training procedure; however, attacks during the testing phase are still~viable.

\subsubsection{Homomorphic~Encryption}
\label{homomorphicencryption}

Homomorphic Encryption (HE) \cite{fontaine2007survey}, is a complex cryptographic protocol, which allows the mathematical computation of encrypted data. The~outcome of the computation is still encrypted. HE is a promising method to protect both the training and testing procedures; however, such an intensive technique's high computational cost is not tolerable in real-world situations. Several HE schemes in the literature propose alterations and evaluations of the method~\cite{gentry2009fully,bost2015machine,zhang2016review,sathya2018review,kairouz2019advances}.

\subsection{Related~Work}
\label{related}

Our work is not another defensive method or technique that mitigates the aforementioned FL attacks, as~seen in Section~\ref{FLattacks}. Hence, a~comparison with defensive techniques such as Knowledge Distillation~\cite{hinton2015distilling,papernot2016distillation}, Anomaly Detection~\cite{song2017machine}, Privacy Engineering~\cite{gurses2016privacy}, Privacy-Preserving Record Linkage~\cite{franke2019scads}, Adversarial Training~\cite{tramer2017ensemble}, ANTIDOTE~\cite{rubinstein2009antidote}, Activation Clustering~\cite{chen2018detecting}, Fine-pruning~\cite{liu2018fine}, STRIP~\cite{gao2019strip}, or~similar, is not comparable and out of the scope of this~paper.

The concern related to the privacy of the stored data has been extensively researched in the literature. Many researchers proposed and presented novel infrastructures and concepts that could partially or fully protect data. However, the~privacy-preservation of critical data such as medical records often is more important than the actual procedure that it has been used, such as the training of a ML algorithm. There are works that presented the use of another emerging technology such as blockchain, which could be combined with ML. In~the work of~\cite{stamatellis2020privacy,papadopoulos2020privacy} the authors' presented infrastructures that could protect certain private data from the stored records and display of other non-private. However, the~feasibility of performing ML in data stored in their blockchain has not been tested and remains an open~question.

Another state-of-the-art technology that is similar to our work is the Private Set Intersection (PSI). Using PSI, participants of an infrastructure can compare the private records they share, without~disclosing them to the other participants~\cite{dachman2009efficient}. There are applications that use PSI for privacy-preserving contact tracing systems and machine learning on vertically partitioned datasets~\cite{angelou2020asymmetric}.

The value of developing an ecosystem and associated governance framework to facilitate the issuance and verification of integrity assured attributes had been considered previously in a different setting~\cite{abramson2020trust}. This work develops user-led requirements for a staff passporting system to reduce the administrative burden placed on healthcare professionals as they interact with different services, employers and educational bodies throughout their careers. These domain-specific systems that digitally define trustworthy entities, policies and information flows have multiple use cases and appear to be a positive indication of the likelihood of broader adoption of the technologies discussed in this~paper.

Achieving privacy-enhancing identity management systems has been a focus of cryptographic research since Chaum published his seminal paper in 1985~\cite{chaum1985security}. The~Hyperledger technology stack used in this work is an implementation of a set of protocols formalised by Camenisch and Lysyanskya~\cite{camenisch2001efficient,camenisch2002signature} and follows a technical architecture closely aligned to the one produced as part of an EU grant ABC4Trust~\cite{bichsel2014d2}. Self-Sovereign Identity (SSI) has popularised the model of issuer/verifier/holder, with~many different projects and implementations building to the emerging standards in this area~\cite{dunphy2018first}, although~not all of these projects use privacy-enhancing cryptography. Furthermore, the~major focus of these systems has been the identification and authentication of individuals by issuing and verifying their credentials within a certain context~\cite{abramson2020trust,wang2020self}.

Our work differentiates from the other approaches since we focus on modelling trust relationships between organisations using the mental model of SSI and the technology stack under development in the Hyperledger foundation. More specifically, in~our proof-of-concept, we establish trusted connections among only authorised participants and then perform FL over secure communication channels. {In our previous work~\cite{abramson2020distributed}, we presented a proof-of-concept that is able to establish trust between the participating parties and perform FL on their data. Thus, to~achieve it, we used the basic messages protocol provided by Hyperledger Aries, encoded the ML model and updates into text format and sent it through DIDComm encrypted channels. However, in~this paper, we have refactored this functionality into libraries that developed mutually within the OpenMined open-source community in PyDentity~\cite{PyDentity}, for~that purpose. Additionally, we have thoroughly presented their communication details in Section~\ref{sectioncommprotocol}, and~have tested their security in Section~\ref{securitytesting}.}

\section{Implementation~Overview}
\label{methodology}

\begin{figure}[H]
\includegraphics[width=0.8\linewidth]{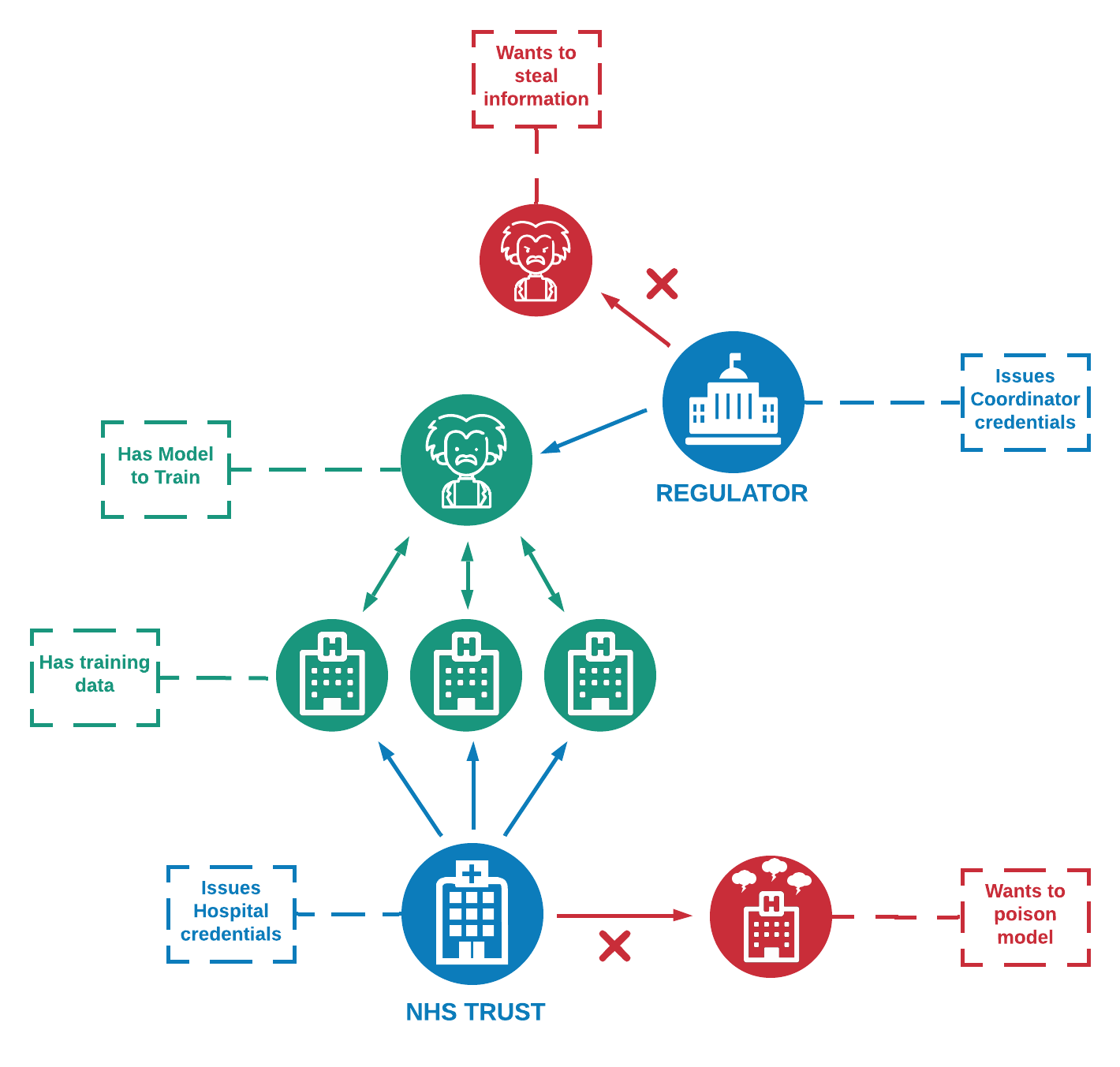}
\caption{ML Healthcare 
Trust Model~\cite{abramson2020distributed}.}
\label{fighla_agents}
\end{figure}

{In our implementation, we used the Hyperledger Aries framework to create a distributed FL architecture~\cite{abramson2020distributed}}. The~communication between the participating entities takes place through the DIDComm transport protocol. We present a healthcare trust model in which each participant is in the form of a Docker container. Our architecture can be seen in Figure~\ref{fighla_agents} and consists of three Hospitals, one Researcher, one NHS Trust that issues Hospitals' credentials, and~a regulatory authority that issues the Researcher's credentials. The~system's technical specifications are as follows: 3.2 GHZ 8th generation Intel Core i7 CPU, with~32 GB RAM and 512 GB SSD. Each Docker container functions as a Hyperledger Aries agent and built using the open-source Hyperledger Aries cloud agent in Python programming language, developed by the Verifiable Organizations Network (VON) team at the Province of British Columbia~\cite{aries_cloud_python}.

\subsection{Establishing~Trust}
\label{sectionestabltrust}

We create a domain-specific trust architecture by using DIDs and VCs issued by trusted participants, as~presented in Section~\ref{methodology}. Furthermore, during~the connection establishment between the Hospitals and the Researcher, they need to follow a mutual authentication procedure, in~which they present their issued credentials as proof that they are legitimate. The~other party can then verify if the received credential has been issued by the public DID of the regulatory authority or the NHS Trust and approve the connection. The~credential schema and the DIDs of the credential issuers are written to a public ledger; we used British Columbia VON's~\cite{Britishcolumbia} development~ledger.

To create our testbed infrastructure, the~steps described in Algorithm~\ref{Algorithm:Trust} followed. After~the authentication is completed, the~Researcher can initiate a FL mechanism, in~which the ML model is being sent encrypted to each Hospital from the list of approved connections sequentially. Each Hospital trains this model using its mental health dataset and sends the model back to the Researcher. Then, the~Researcher validates the trained model using its validation dataset and sends it to the next Hospital. The~presented procedure continues until all the participants train the ML model. At~the end of the training, the~Researcher holds a ML model trained by multiple Hospitals that validated using its validation dataset to calculate the model's accuracy and loss. The~model's parameters and updates are being sent using the DIDComm transport protocol, the~security and performance evaluation of it have been presented and discussed in Section~\ref{evaluation}.

\subsection{Communication~Protocol}
\label{sectioncommprotocol}

\begin{figure}[H]
\widefigure
\includegraphics[width=1.0\linewidth]{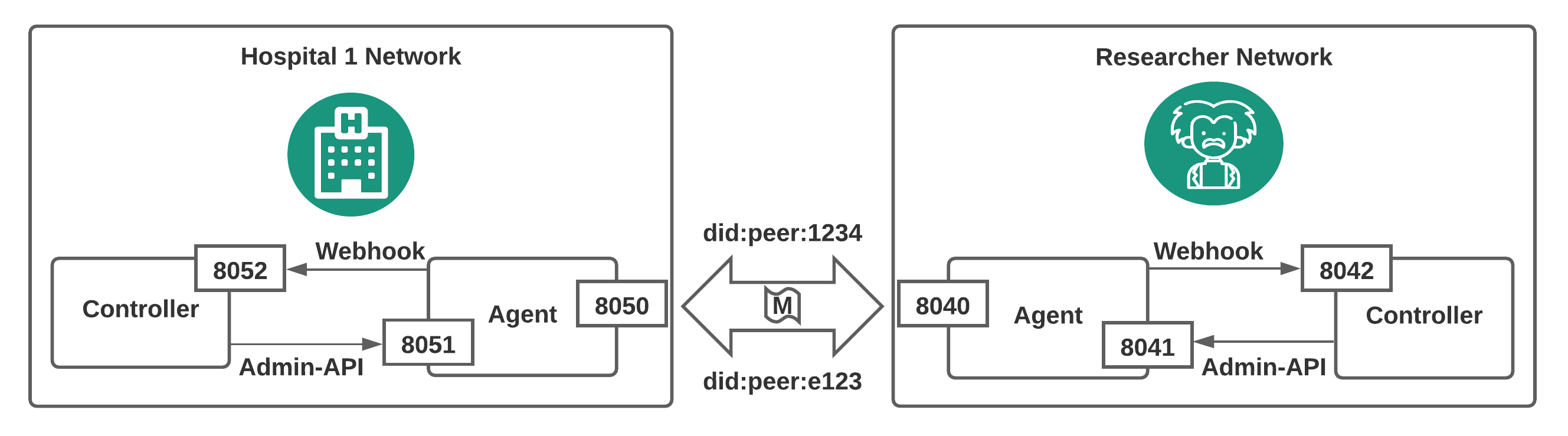}
\caption{Networking communication architecture.}
\label{fig:agent_networking}
\end{figure}

As presented in Section~\ref{methodology} and Figure~\ref{fighla_agents}, our implementation consists of three Hospitals, one Researcher that coordinates the training procedure, an~NHS Trust and a Regulator that issue credentials for the Hospitals and the Researcher accordingly. Each participating entity is configured as a Docker container instead of a real-world situation in which each participant would run on their own network. Communication between entities only happens using the DIDComm protocol adding an authenticated encryption layer (see Algorithm \ref{Algorithm:DID}) on top of the underlying transport protocol, in~this case, Hypertext Transfer Protocol (HTTP) at a specified public port. Internally the entity can be represented as a controller and an agent; the controller sends HTTP requests to the agent defined by the admin-API. Received requests act as commands often resulting in the agent sending a DIDComm protocol message to an external agent, for~example issuing a credential. Agents that receive a message from another entity post a webhook internally over HTTP, allowing the controller to respond appropriately. Note this can include requesting the agent to send further messages in reply. More details can be seen in Figure~\ref{fig:agent_networking} and Table~\ref{tab:networkdetails}.

\begin{algorithm}[H]
\caption{Establishing Trusted Connections~\cite{abramson2020distributed}}
\label{Algorithm:Trust}
\begin{spacing}{1.2}
\begin{algorithmic}[1]
\STATE \textit{Researcher} agent exchanges DIDs with the \textit{Regulator} agent to establish a DIDComm channel.
\STATE \textit{Regulator} offers an \textit{Audited Researcher-Coordinator} credential over this channel.
\STATE \textit{Researcher} accepts and stores the credential in their wallet.
\FOR{each \textit{Hospital} agent}
\STATE Initiate DID Exchange with \textit{NHS Trust} agent to establish DIDComm channel.
\STATE \textit{NHS Trust} offers \textit{Verified Hospital} credentials over DIDComm.
\STATE \textit{Hospital} accepts and stores the credential.
\ENDFOR
\FOR{each \textit{Hospital} agent}
\STATE \textit{Hospital} initiates DID Exchange with \textit{Researcher} to establish DIDComm channel.
\STATE \textit{Researcher} requests proof of \textit{Verified Hospital} credential issued and signed by the \textit{NHS Trust}.
\STATE \textit{Hospitals} generate a valid proof from their \textit{Verified Hospital} credential and respond to the \textit{Researcher}.
\STATE \textit{Researcher} verifies the proof by first checking the DID against the known DID they have stored for the \textit{NHS Trust}, then \textit{resolve} the DID to locate the keys and verify the signature.
\IF{\textit{Hospitals} can prove they have a valid \textit{Verified Hospital} credential}
\STATE \textit{Researcher} adds the connection identifier to their list of \textit{Trusted Connections}.
\ENDIF
\STATE \textit{Hospital} requests proof of \textit{Audited Researcher} credential from the \textit{Researcher}.
\STATE \textit{Researcher} uses \textit{Audited Researcher} credential to generate a valid proof and responds.
\STATE \textit{Hospital} verifies the proof, by~checking the signature and DID of the Issuer.
\IF{\textit{Researcher} produces a valid proof of \textit{Audited Researcher}}
\STATE \textit{Hospital} saves connection identifier as a trusted connection.
\ENDIF
\ENDFOR
\end{algorithmic}
\end{spacing}
\end{algorithm}
\unskip

\end{paracol}
\nointerlineskip
\begin{paracol}{2}
\switchcolumn

\unskip

\begin{specialtable}[H]
\caption{Participating entities communication details.}
\setlength{\cellWidtha}{\columnwidth/4-2\tabcolsep+0.0in}
\setlength{\cellWidthb}{\columnwidth/4-2\tabcolsep+0.0in}
\setlength{\cellWidthc}{\columnwidth/4-2\tabcolsep+0.0in}
\setlength{\cellWidthd}{\columnwidth/4-2\tabcolsep+0.0in}
\scalebox{1}[1]{\begin{tabularx}{\columnwidth}{>{\PreserveBackslash\centering}m{\cellWidtha}>{\PreserveBackslash\centering}m{\cellWidthb}>{\PreserveBackslash\centering}m{\cellWidthc}>{\PreserveBackslash\centering}m{\cellWidthd}}
\toprule
\textbf{Name} & \textbf{HTTP Port} & \textbf{Admin-API Port} & \textbf{Webhook Port} \\ \midrule
Hospital 1 & 8050 & 8051 & 8052 \\ \midrule
Hospital 2 & 8060 & 8061 & 8062 \\ \midrule
Hospital 3 & 8070 & 8071 & 8072 \\ \midrule
Researcher & 8040 & 8041 & 8042 \\ \midrule
NHS Trust & 8020 & 8021 & 8022 \\ \midrule
Regulator & 8030 & 8031 & 8032 \\ \bottomrule
\end{tabularx}}
\label{tab:networkdetails}
\end{specialtable}
\unskip

\subsection{Federated Learning~Procedure}
\label{sectionvanillafl}

\textls[-15]{{The FL procedure described in our proof-of-concept is in its most basic form, in~which the model and the updates are being sent sequentially to each trusted \mbox{connection~\cite{nishio2019client,shoham2019overcoming,kopparapu2020fedfmc,konevcny2016federated}}. In~a real-world scenario, this FL process would happen simultaneously, and~the model updates would be sent to a secure aggregator to perform a Federated Averaging method~\cite{konevcny2016federated} to improve the security of the system further}. Before~the training, the~Researcher holds a ML model and a validation dataset, and~each Hospital holds its own training dataset. The~datasets are from an open-source mental health survey that ``that measures attitudes towards mental health and frequency of mental health disorders in the tech workplace''~\cite{Mentalhealthdataset}, which are pre-processed into appropriate training data and validation data; the original dataset split into four partitions, three training datasets, one for each Hospital and one validation dataset for the~Researcher.}

Furthermore, we evaluated our infrastructure's performance related to the model's accuracy and measured the required resources. Our FL workflow can be seen in \mbox{Algorithm~\ref{Algorithm:FL}}. The~focus of this paper is to demonstrate that FL is applicable over Hyperledger Aries agents through the DIDComm protocol in a trusted architecture scenario. Therefore, the~ML procedure, classification and parameter-tuning are out of the scope of this~paper.

However, the~combination of these two emerging fields, private identities and FL, allowed us to mitigate a few existing FL limitations caused by the training participants' lack of trust. Specifically, these were: (1) Training provided by a malevolent Hospital to corrupt the ML model's accuracy, and~(2) Malicious models being sent to legitimate Hospitals to leak information about the training~data.

\begin{algorithm}[H]
\caption{Our Federated Learning workflow~\cite{abramson2020distributed}}
\label{Algorithm:FL}
\begin{spacing}{1.2}
\begin{algorithmic}[1]
\STATE \textit{Researcher} has \textit{validation} data and a \textit{ML model}, \textit{Hospitals} have \textit{training} data.
\WHILE{\textit{Hospitals} have not trained their \textit{training} data}
\STATE \textit{Researcher} benchmarks the \textit{model's} performance against \textit{validation} data and sends the \textit{model} to the next \textit{Hospital}.
\STATE \textit{Hospital} trains the \textit{model} with their data and then sends the resulting \textit{model} back to the \textit{Researcher}.
\ENDWHILE
\STATE \textit{Researcher} benchmarks the final \textit{model} against \textit{validation} data.
\end{algorithmic}
\end{spacing}
\end{algorithm}

\section{Evaluation}
\label{evaluation}

\subsection{Security~Evaluation}
\label{sectionseceval}

As presented in our implementation in Section~\ref{methodology}, our testbed infrastructure achieves a domain-specific trust framework using verifiable credentials. Hence, the~training process involves only authenticated Hospitals and Researchers that communicate through encrypted channels. Our work does not prevent the aforementioned attacks from happening; however, it minimises the possibility of occurring by establishing a trust framework among the participants. Malicious entities could be checked on their registration to the system and removed on ill~behaviour.

{A potential threat in our test environment is the possibility of the participants' computer systems getting compromised. In~such scenarios, the~trusted credential issuers could create legitimate credentials to malicious participants, or~a compromised hospital could corrupt the ML training; both scenarios lead to a malicious participant controlling a valid VC for the infrastructure. Another concern is the possibility that a compromised participant may try a Distributed Denial of Service (DDoS) attack~\cite{lau2000distributed}, which can be mitigated by setting a timeout process within each participant, after~several unsuccessful invitations. Several cybersecurity procedures could be in-place within the participants' computer systems that make security concerns and breaches unlikely. OWASP provides several secure practices and guidelines in order to mitigate cybersecurity threats~\cite{owasp201810}; hence, further defensive mechanisms could be used to extend further the security of the system, such as Intrusion Detection and Prevention Systems (IDPS) \cite{Modelvuln2019}. However, this type of attack is out of the scope of this paper.}

To evaluate our proof-of-concept's security, we created malicious agents that attempt to take part in the ML procedure by connecting to one of the trusted credential issuers. Any agent without the appropriate VCs, either a verified Hospital or an audited Researcher credential, could not establish an authenticated channel with the other party, as~seen in Figure~\ref{fighla_agents}. The~unauthorised connection requests and the self-signed VC are automatically being rejected. The~reason is because they had not been signed by a trusted authority whose DID was known by the entity requesting the proof. The~mechanism of the mutual authentication of the VC between participants is not domain-specific to ML and can be expanded to any context~\cite{abramson2020distributed}.

However, our paper is focused on the trust establishment and FL in a distributed DID-based healthcare ecosystem. It is assumed that there is a governance-oriented framework in which the key-stakeholders have a DID written to an integrity-assured blockchain ledger. Identifying the appropriate DIDs and the participating entities related to a particular architecture is out-of-scope of this paper. This paper explores how peer DID connections facilitate participation in the established healthcare ecosystem. Another platform could be developed for the secure distribution of the DIDs between the participating agents~\cite{abramson2020distributed}.

\subsubsection{Security~Testing}
\label{securitytesting}

In our implementation, the~ML model and its updates are being sent to the Hospitals and the Researcher is using the DIDComm messaging protocol~\cite{didcomm_dif}. To verify these communication channels were encrypted, we used network packet sniffers such as the Wireshark and Tcpdump~\cite{goyal2017comparative}, to~capture the traffic during the training procedure. As~presented in Section~\ref{dockers}, since the participating entities in our implementation take the form of Docker containers, the~captured traffic obtained from a virtual network card in the host machine~\cite{martin2018docker}. During~the security testing of our implementation, we observed that when the participating entities connect and provide their proofs to the other party in order to authenticate, information such as the name of the participant, its DID and the provided proof are encrypted. Only in case that the Hyperledger Aries agent and controller reside within the same Docker container, then the information related to the connection establishment sent internally in plain \textit{.json} format (Appendix~\ref{appsecuritytesting}, Figure~\ref{figproofs}). However, this finding is irrelevant in production environments since each participant, its controller and agent would be physical machines or private networks and not Docker containers. We demonstrated this separation of controllers and agents, by~executing each one in their own respective Jupyter Notebook~\cite{kluyver2016jupyter}, as~described in Section~\ref{sectioncommprotocol}.

Furthermore, other critical findings were observed during the training procedure. All the traffic was fully encrypted between all the parties, in~each stage of the FL training (Appendix~\ref{appsecuritytesting}, Figure~\ref{figencrypted}). We used unsuccessfully various cybersecurity tools such as the Government Communications Headquarters (GCHQ) CyberChef \cite{Cyberchef} in order to reverse the encrypted content and obtain some information about the training data or the ML model. That is re-assuring since the traffic related to the training procedure may contain information about the sensitive underlying training~data.

\subsection{Performance~Evaluation}
\label{performanceeval}

Performance evaluation metrics for each host were recorded during the operation of our workflow. Figure~\ref{fig:networkoutput}a) shows the CPU usage of each agent involved in the learning workflow. The~CPU usage of the Researcher raises each time it sends the model to the Hospitals, and~the CPU usage of the Hospitals raises when they train the model with their private data. This result is expected and follows the execution of \mbox{Algorithm \ref{Algorithm:FL}} successfully. The~memory and network bandwidth follow a similar pattern, as~it is illustrated in \mbox{Figure~\ref{fig:networkoutput}b--d)}. The~main difference is that since the Researcher averages and validates each model against the training dataset every time, the~memory and network bandwidth increase over time. In~these metrics, the~ML training procedure transmitted through DIDComm protocols but does not use the designed federated learning~libraries.

In Figure~\ref{fig:cpuFLcomparison}, we compared the FL training performance with and without the DIDComm protocol. Both architectures are identical, with~their only difference in~\mbox{Figure~\ref{fig:cpuFLcomparison}a)} where the Researcher sends the ML model to each Hospital sequentially through the DIDComm protocol, opposed to Figure~\ref{fig:cpuFLcomparison}b), in~which the DIDComm protocol is not used, and~the ML model is shared among the Hospitals to train it sequentially. We did not plot the memory and network metrics for this experiment since they follow the same pattern with negligible differences as in Figure~\ref{fig:networkoutput}.

{Our work aims to demonstrate that since the proposed trust framework is distributed, it is possible to establish a FL workflow. Therefore, we do not focus on improving this FL process and tuning the hyperparameters for more reliable predictions, apart from developing the FL libraries designed for this purpose. The~FL training procedure consists of the following hyperparameters: learning rate of 0.01 for 10 training epochs using one-third of the training dataset in batches of size 8. Moreover, we present the ML model's confusion matrix using the Researcher's validation data after each federated training batch, as~shown in Tables~\ref{table:ConfMat} and~\ref{table:ConfMatRelu}. That confirms that our ML model was successfully trained at each stage using our distributed mental health dataset~\cite{Mentalhealthdataset}. To~calculate the model's accuracy, Equation~(\ref{eqn:accuracy}) has been used. The~two tables provide a comparison between two different activation functions. In~Table~\ref{table:ConfMat}, the~Sigmoid linear activation function has been used, as~opposed to Table~\ref{table:ConfMatRelu}, in~which we implemented the Rectified Linear activation function (ReLu) \cite{agostinelli2014learning,nwankpa2018activation}.}
\vspace{-6pt}

\begin{specialtable}[H]
\caption{{Classifier's accuracy without hyperparameters' optimisation over training batches using Sigmoid activation function on the original federated learning architecture of~\cite{abramson2020distributed}.}}
\setlength{\cellWidtha}{\columnwidth/5-2\tabcolsep+0.4in}
\setlength{\cellWidthb}{\columnwidth/5-2\tabcolsep-0.1in}
\setlength{\cellWidthc}{\columnwidth/5-2\tabcolsep-0.1in}
\setlength{\cellWidthd}{\columnwidth/5-2\tabcolsep-0.1in}
\setlength{\cellWidthe}{\columnwidth/5-2\tabcolsep-0.1in}
\scalebox{1}[1]{\begin{tabularx}{\columnwidth}{>{\PreserveBackslash\centering}m{\cellWidtha}>{\PreserveBackslash\centering}m{\cellWidthb}>{\PreserveBackslash\centering}m{\cellWidthc}>{\PreserveBackslash\centering}m{\cellWidthd}>{\PreserveBackslash\centering}m{\cellWidthe}}
\toprule
\textbf{Batch} & \textbf{0} & \textbf{1} & \textbf{2} & \textbf{3} \\ 
\midrule 
True Positives & 0 & 109 & 120 & 134\\
False Positives & 0 & 30 & 37 & 41 \\
True Negatives & 114 & 84 & 77 & 73\\
False Negatives & 144 & 35 & 24 & 10\\
\midrule
Accuracy & 44.1\% 
& 74.8\% & 76.3\% & 80.2\%\\
\bottomrule
\end{tabularx}}
\label{table:ConfMat}
\end{specialtable}
\unskip

\begin{specialtable}[H]
\caption{{Classifier's accuracy without hyperparameters' optimisation over training batches using ReLu activation function}.}
\setlength{\cellWidtha}{\columnwidth/5-2\tabcolsep+0.4in}
\setlength{\cellWidthb}{\columnwidth/5-2\tabcolsep-0.1in}
\setlength{\cellWidthc}{\columnwidth/5-2\tabcolsep-0.1in}
\setlength{\cellWidthd}{\columnwidth/5-2\tabcolsep-0.1in}
\setlength{\cellWidthe}{\columnwidth/5-2\tabcolsep-0.1in}
\scalebox{1}[1]{\begin{tabularx}{\columnwidth}{>{\PreserveBackslash\centering}m{\cellWidtha}>{\PreserveBackslash\centering}m{\cellWidthb}>{\PreserveBackslash\centering}m{\cellWidthc}>{\PreserveBackslash\centering}m{\cellWidthd}>{\PreserveBackslash\centering}m{\cellWidthe}}
\toprule
\textbf{Batch} & \textbf{0} & \textbf{1} & \textbf{2} & \textbf{3} \\ 
\midrule 
True Positives & 144 & 121 & 121 & 108\\
False Positives & 0 & 23 & 23 & 36 \\
True Negatives & 114 & 34 & 33 & 39\\
False Negatives & 0 & 80 & 81 & 75\\
\midrule
Accuracy & 100\% & 60\% & 59,6\% & 57\%\\
\bottomrule
\end{tabularx}}
\label{table:ConfMatRelu}
\end{specialtable}

\begin{figure}[H]
\subfloat[CPU Usage (\%) during workflow]{{\includegraphics[width=0.45\linewidth]{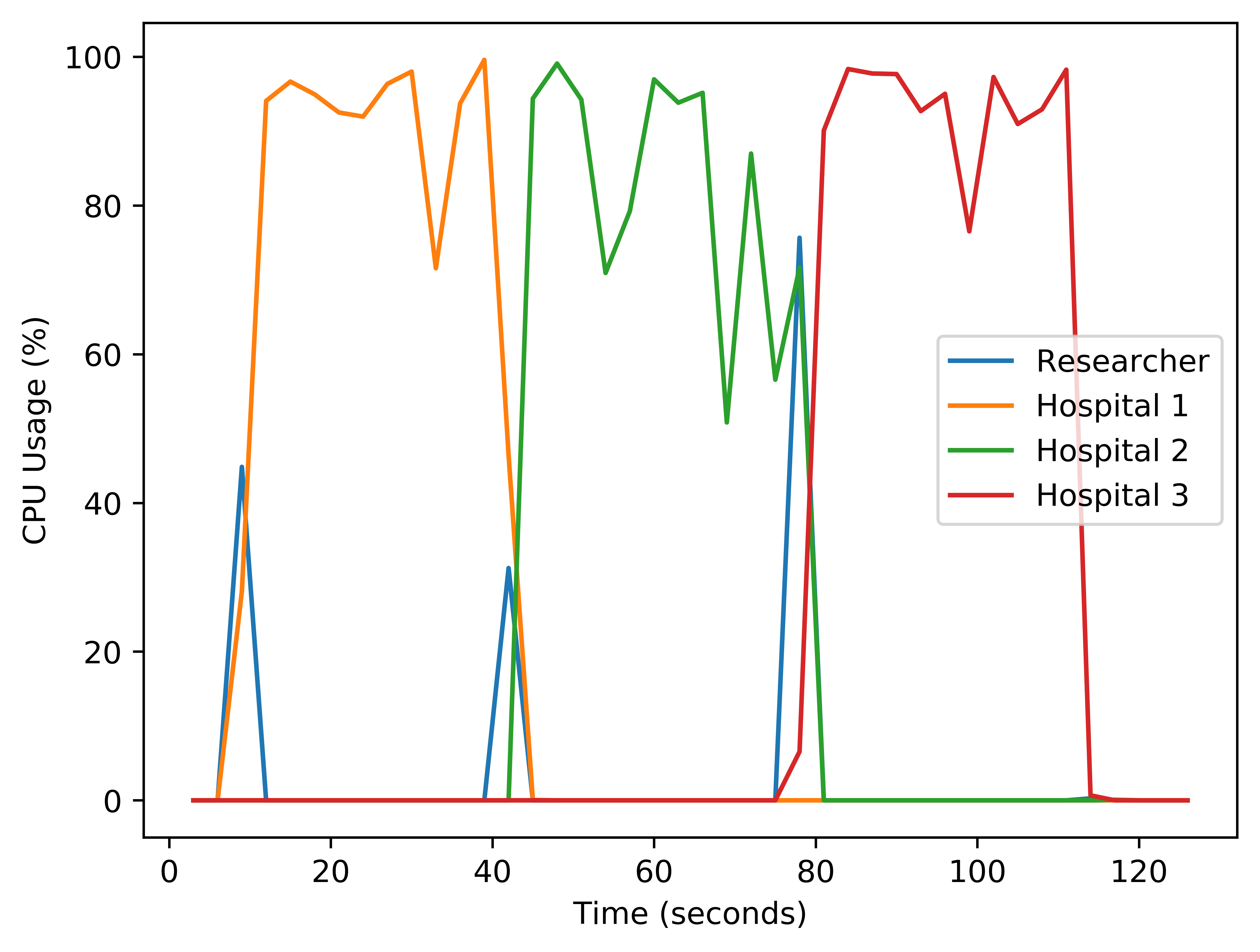} }}
\qquad
\subfloat[Memory Usage (\%) during workflow]{{\includegraphics[width=0.45\linewidth]{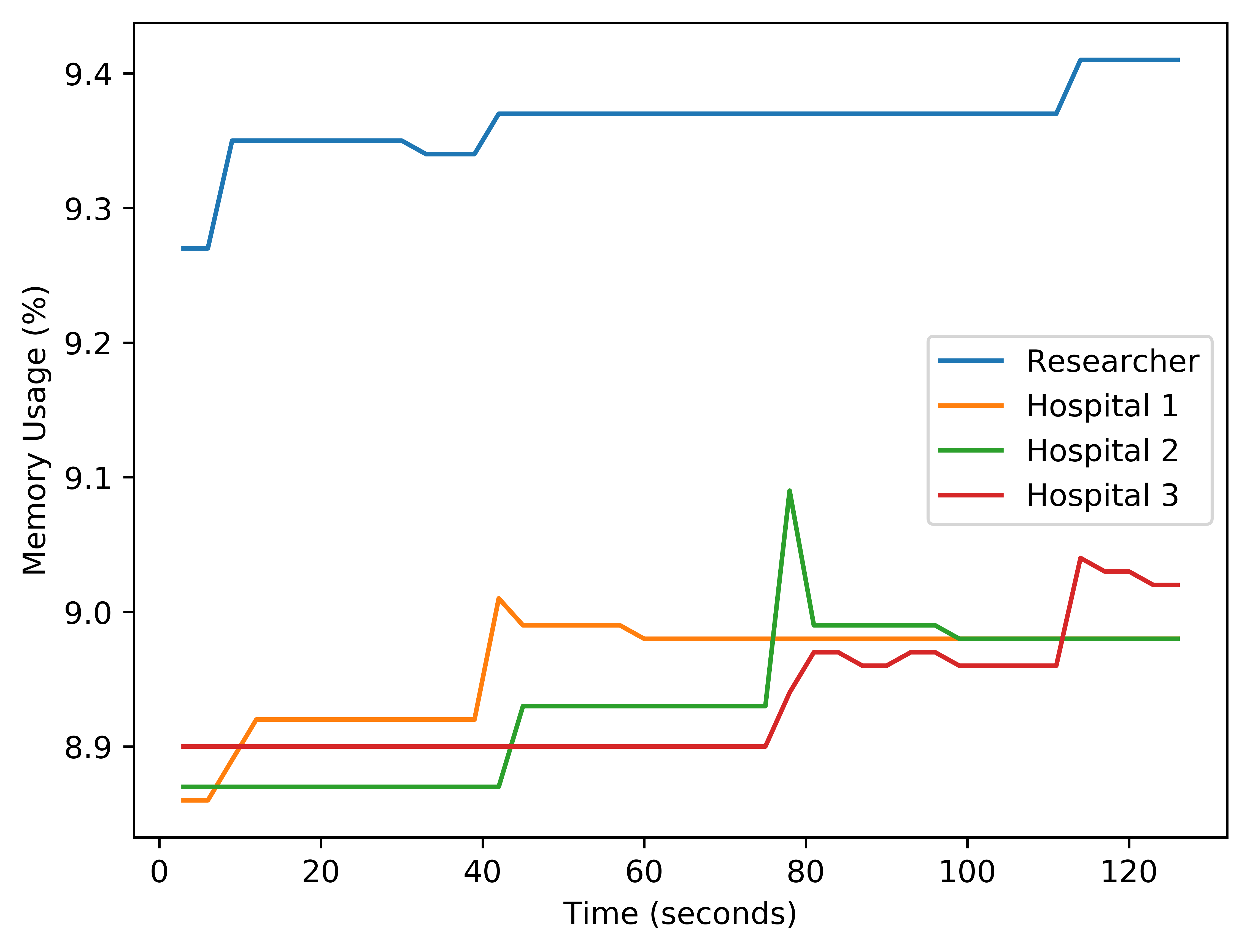} }}
\qquad
\subfloat[Network Input (kB) during workflow]{{\includegraphics[width=0.45\linewidth]{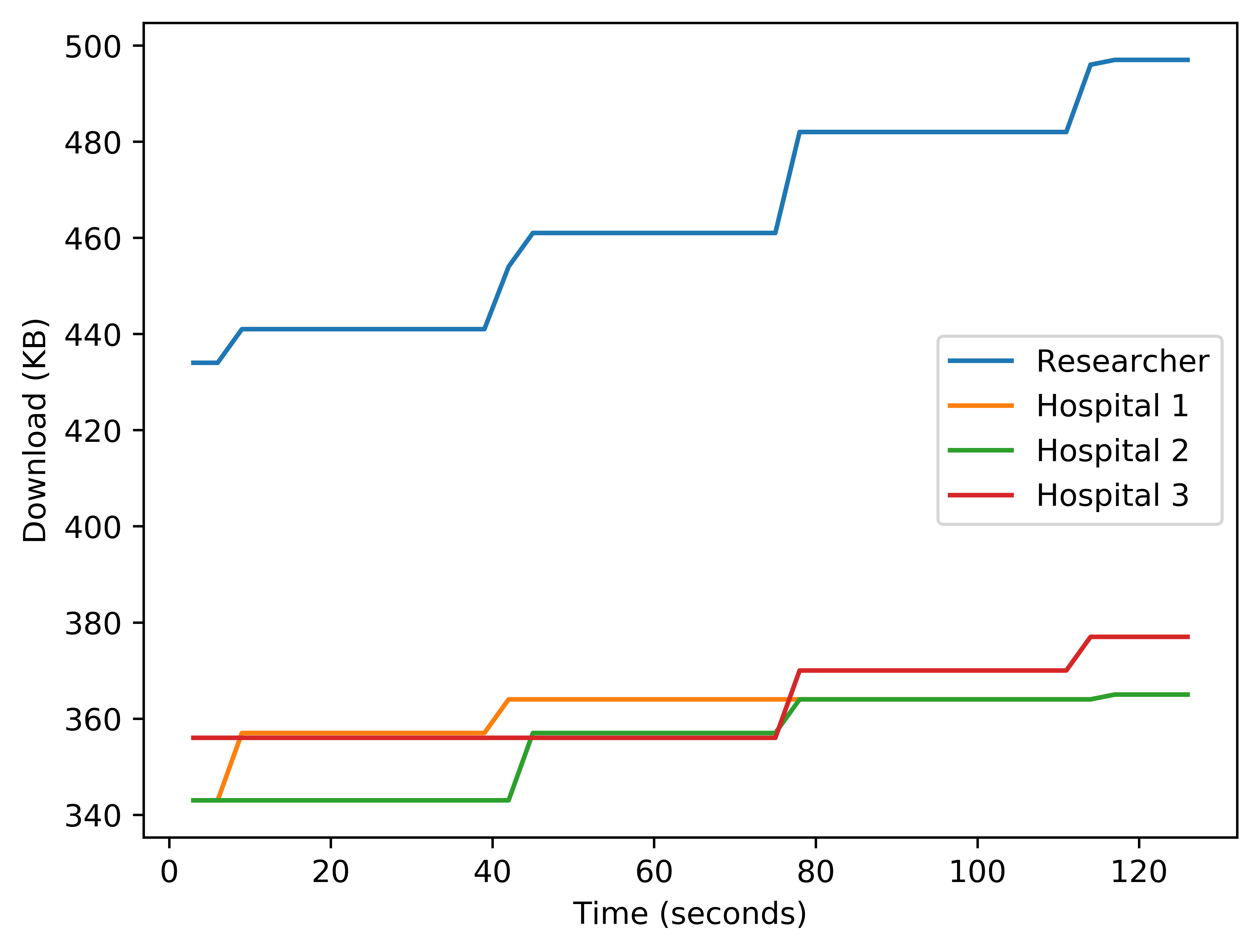} }}
\qquad
\subfloat[Network Output (kB) during workflow]{{\includegraphics[width=0.45\linewidth]{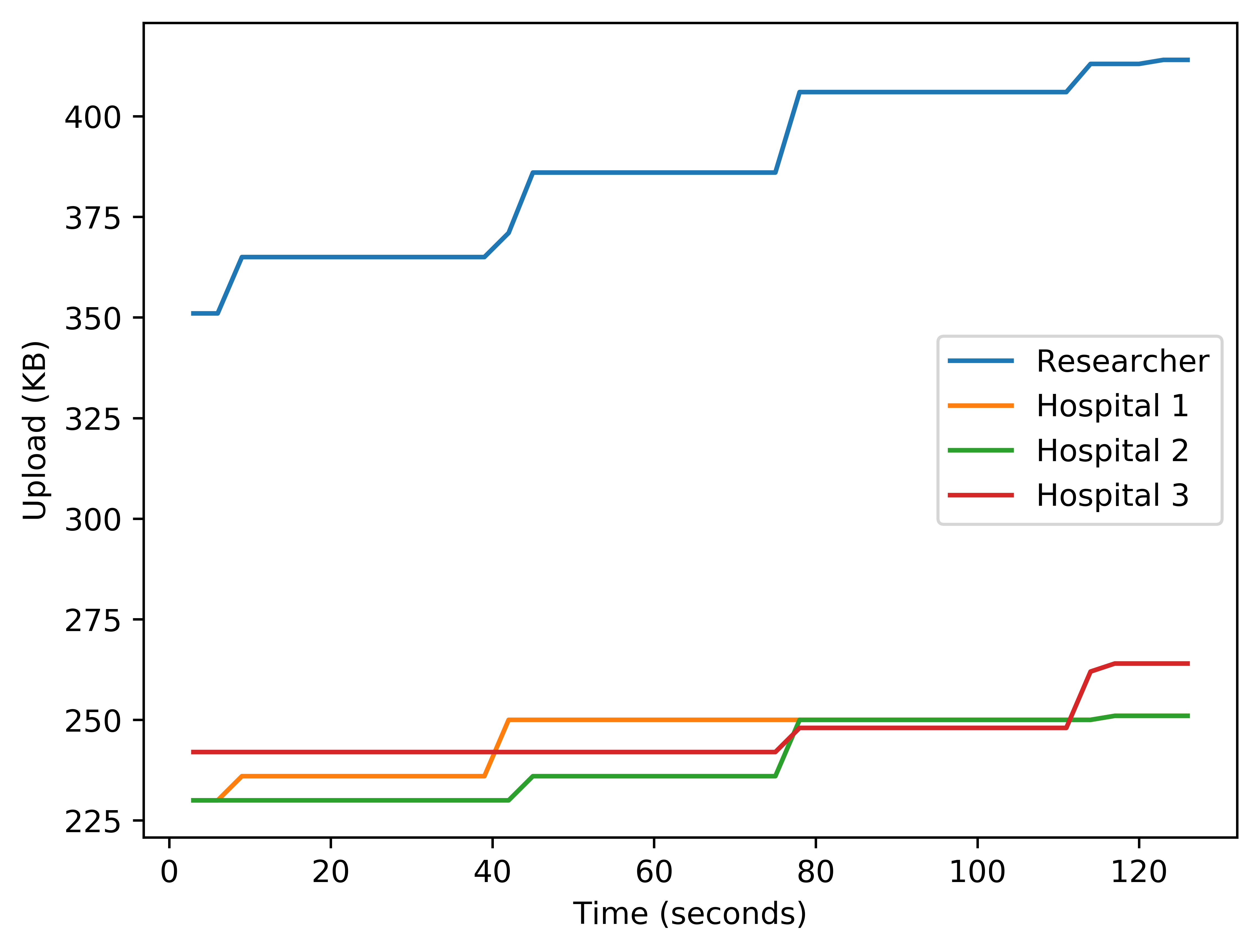} }}
\caption{CPU, Memory usage and Network use of Docker container agents during workflow using the original federated learning architecture~\cite{abramson2020distributed}.}
\label{fig:networkoutput}%
\end{figure}


Moreover, we also compared the FL procedure's accuracy when the ML model is being transmitted through and without the DIDComm protocol, and~presented the results in Tables~\ref{table:ConfMat_aries-FL} and~\ref{table:ConfMat_FL}.

\begin{specialtable}[H]
\caption{Classifier's accuracy without hyperparameters' optimisation over training batches through the DIDComm protocol.} 
\setlength{\cellWidtha}{\columnwidth/5-2\tabcolsep+0.4in}
\setlength{\cellWidthb}{\columnwidth/5-2\tabcolsep-0.1in}
\setlength{\cellWidthc}{\columnwidth/5-2\tabcolsep-0.1in}
\setlength{\cellWidthd}{\columnwidth/5-2\tabcolsep-0.1in}
\setlength{\cellWidthe}{\columnwidth/5-2\tabcolsep-0.1in}
\scalebox{1}[1]{\begin{tabularx}{\columnwidth}{>{\PreserveBackslash\centering}m{\cellWidtha}>{\PreserveBackslash\centering}m{\cellWidthb}>{\PreserveBackslash\centering}m{\cellWidthc}>{\PreserveBackslash\centering}m{\cellWidthd}>{\PreserveBackslash\centering}m{\cellWidthe}}
\toprule
\textbf{Batch} & \textbf{0} & \textbf{1} & \textbf{2} & \textbf{3} \\ 
\midrule 
True Positives & 0 & 115 & 120 & 135\\
False Positives & 0 & 29 & 24 & 9\\
True Negatives & 113 & 30 & 39 & 44\\
False Negatives & 145 & 84 & 75 & 70\\
\midrule
Accuracy & 43.7\% & 56.2\% & 61.6\% & 69.3\%\\
\bottomrule 
\end{tabularx}}
\label{table:ConfMat_aries-FL} 
\end{specialtable}
\unskip

\begin{specialtable}[H]
\caption{Classifier's accuracy without hyperparameters' optimisation over training batches without the DIDComm protocol.} 
\setlength{\cellWidtha}{\columnwidth/5-2\tabcolsep+0.4in}
\setlength{\cellWidthb}{\columnwidth/5-2\tabcolsep-0.1in}
\setlength{\cellWidthc}{\columnwidth/5-2\tabcolsep-0.1in}
\setlength{\cellWidthd}{\columnwidth/5-2\tabcolsep-0.1in}
\setlength{\cellWidthe}{\columnwidth/5-2\tabcolsep-0.1in}
\scalebox{1}[1]{\begin{tabularx}{\columnwidth}{>{\PreserveBackslash\centering}m{\cellWidtha}>{\PreserveBackslash\centering}m{\cellWidthb}>{\PreserveBackslash\centering}m{\cellWidthc}>{\PreserveBackslash\centering}m{\cellWidthd}>{\PreserveBackslash\centering}m{\cellWidthe}}
\toprule
\textbf{Batch} & \textbf{0} & \textbf{1} & \textbf{2} & \textbf{3} \\ 
\midrule 
True Positives & 0 & 113 & 120 & 133\\
False Positives & 0 & 31 & 24 & 11 \\
True Negatives & 116 & 35 & 43 & 45\\
False Negatives & 142 & 79 & 71 & 69\\
\midrule
Accuracy & 44.9\% & 57.3\% & 63.1\% & 69\%\\
\bottomrule 
\end{tabularx}}
\label{table:ConfMat_FL} 
\end{specialtable}

\begin{figure}[H]
\subfloat[CPU Usage (\%) during workflow and transmission of the model through the DIDComm protocol]{{\includegraphics[width=0.57\linewidth]{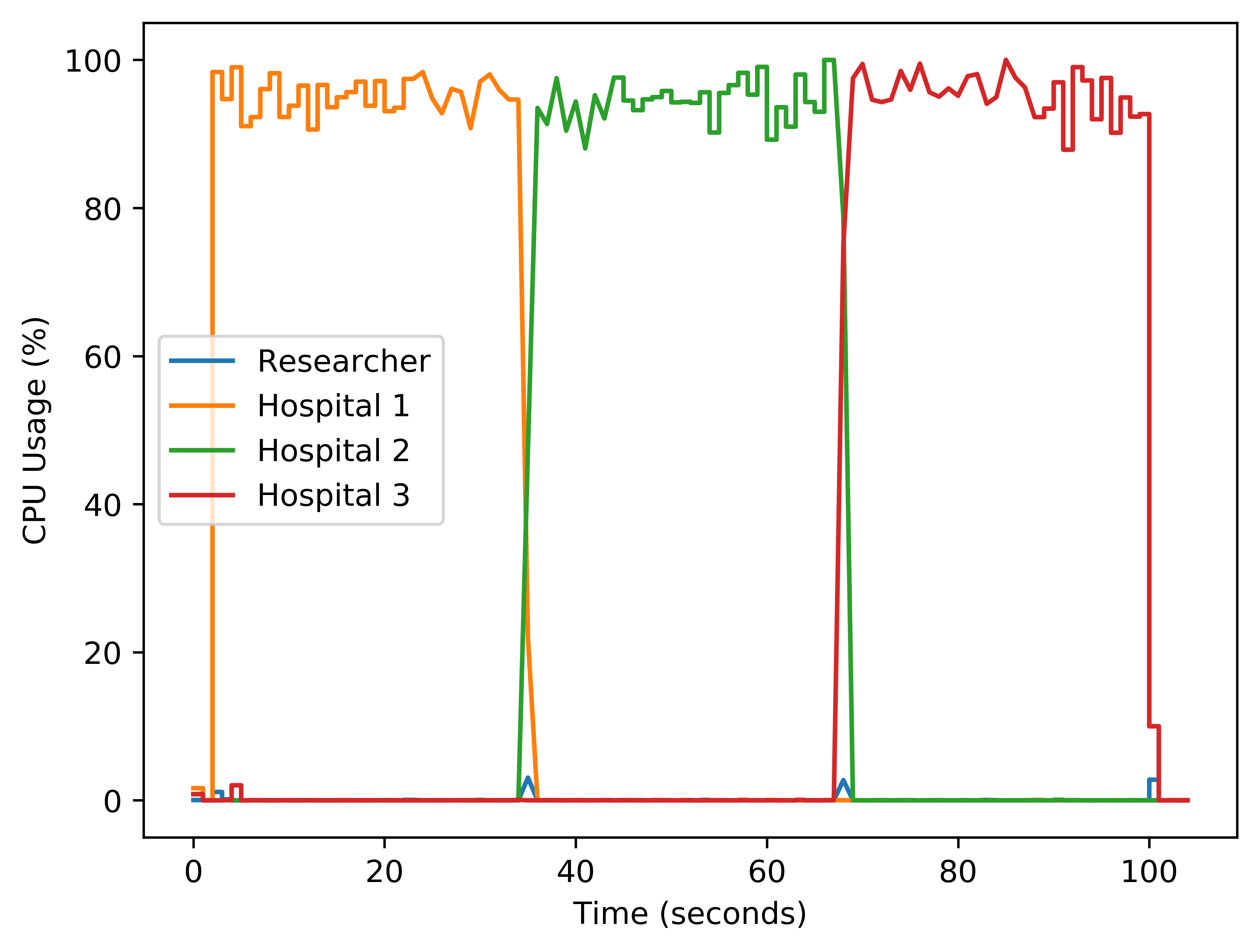} }}\\
\subfloat[CPU (\%) during workflow without the use of the DIDComm protocol]{{\includegraphics[width=0.57\linewidth]{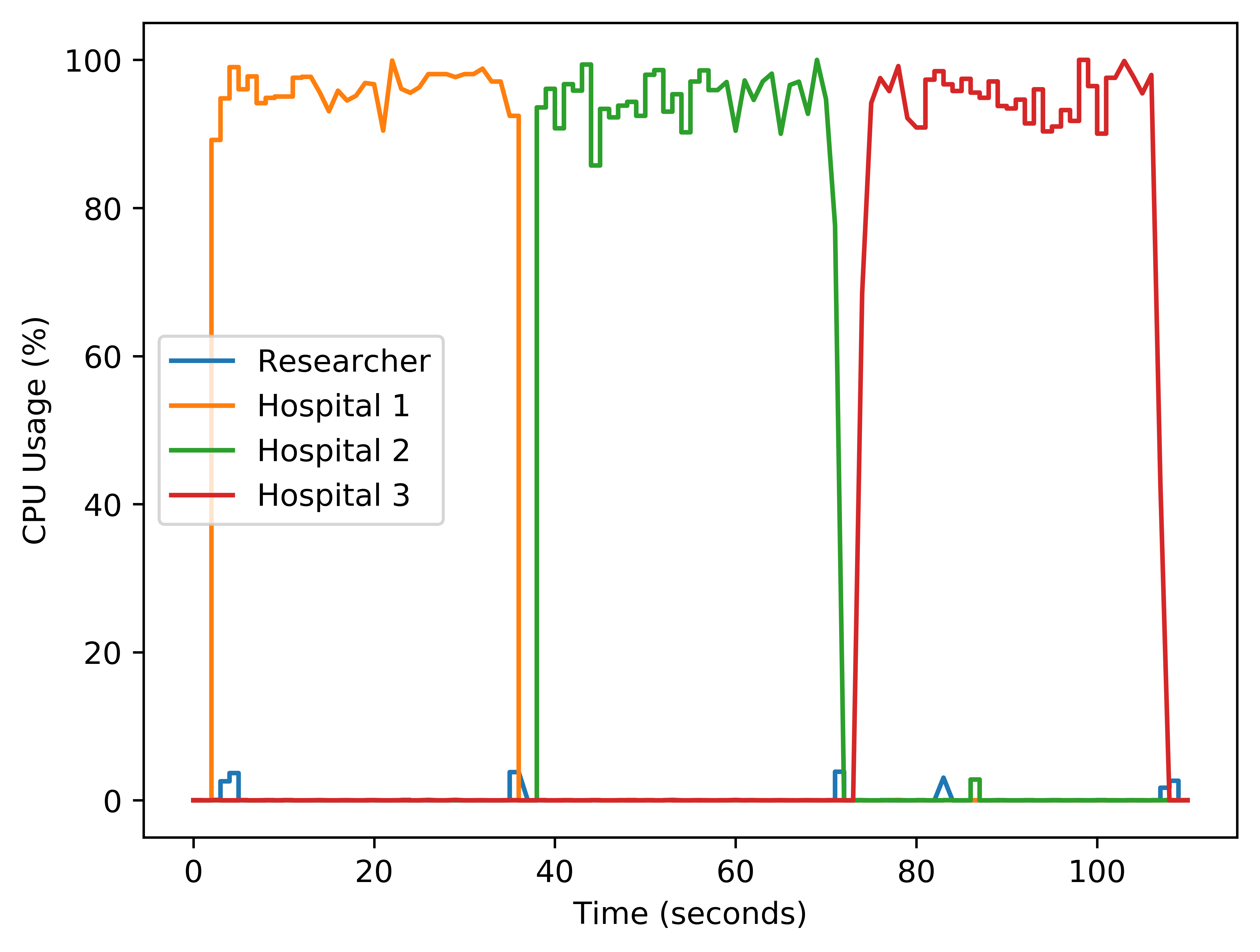} }}
\caption{CPU Usage comparison of Docker containers during workflow using our novel federated learning libraries.}
\label{fig:cpuFLcomparison}%
\end{figure}

\section{Conclusions and Future~Work}
\label{conclusion}

In this paper, we extended our previous work~\cite{abramson2020distributed} by merging the privacy-preserving ML field with VCs and DIDs while addressing trust concerns within the data industry. These areas focus on people's security, privacy and especially on the protection of their sensitive data. In~our work, we presented a trusted FL process for a mental health dataset distributed among hospitals. We proved that it is possible to use the established secure channels to obtain a digitally signed contract for ML training or manage pointer communications on remote data~\cite{Ryffel2018}.

{This extension of our previous work~\cite{abramson2020distributed} retains the same high-level architecture of the participating entities, but~the proof-of-concept is a complete refactor of our experimental setup. More specifically, as~described in Section~\ref{securitytesting}, each participant's controller and agent entities are separated into their own isolated Docker containers. That separation is adjacent to a real-world scenario in which each controller and agent reside in different systems (Appendix~\ref{appsecuritytesting}). Furthermore, in~our technical codebase, we now use our novel libraries written in Python programming language and demonstrated thoroughly
using Jupyter notebooks. We further performed an extensive security and performance evaluation in each stage of our proposed infrastructure, which was lacking in our previous work, and~our findings are presented in Section~\ref{evaluation}. The~performance metrics identified that the performance of our trusted FL procedure and the accuracy of the ML model are similar, albeit the model is transmitted through the encrypted DIDcomm protocol. Additionally, using our designed FL libraries, the~ML training process completes faster. It should be noted that there are no conflicts with other defensive methods and techniques, and~they could be incorporated into our framework and libraries.}

While FL is vulnerable to attacks as described in Section~\ref{sectionseceval}, the~purpose of this work is to develop a proof-of-concept for the demonstration that distributed ML can be achieved through the same encrypted communication channels used to establish domain-specific trust. We exhibited how this distributed trust framework could be used by other fields and not FL explicitly. This will allow the application of the trust framework to a wide range of privacy-preserving workflows. Additionally, it allows us to enforce trust, mitigating FL attacks using differentially private training mechanisms~\cite{dwork2011differential,abadi2016deep,mironov2017renyi}. Various techniques can be incorporated to our framework in order to train a differentially private model; such as Opacus~\cite{Opacus}, PyDP~\cite{PyDP}, PyVacy~\cite{PyVacy} and LATENT~\cite{chamikara2019local}. To~reduce the model stealing and training data inference risks, the~SMPC can be leveraged to split data and model parameters into shares~\cite{lindell2005secure}.

{Our proof-of-concept detailed the established architecture between three hospitals, a~researcher, a~hospital trust and a regulatory authority. Firstly, the~hospitals and the researcher need to obtain a VC from their corresponding trust or regulatory authority, and~then follow a mutual authentication process in order to exchange information. Further, the~researcher instantiates a basic FL procedure between only the authenticated and trusted hospitals, we refer to this process as Vanilla FL, and~then transmits the ML model through the encrypted communication channels using Hyperledger Aries framework. Each hospital receives the model, trains it using their private dataset and sends it back to the researcher. The~researcher validates the trained model using its validation dataset to calculate its accuracy. One of the limitations of this work is that the presented Vanilla FL process acts only as a proof-of-concept to demonstrate that FL is possible through the encrypted DIDComm channels. However, to~incorporate it in a production environment, it should be extended and introduce a secure aggregator entity, placed in-between the researcher and the hospitals that would act as a mediator of the ML model and updates. In~that production environment, the~researcher entity would simultaneously send the ML model to all the authorised participants and not have a validation dataset. This is a crucial future improvement we need to undertake to help the research community further. Another potential limitation of our work is training a large-scale convolutional neural network, which left as out-of-scope, but~needs to be tested.}

Future work also includes integrating the Hyperledger Aries communication protocols, which enables the trust model demonstrated in this work, into~an existing framework for facilitating distributed learning within the OpenMined open-source organisation such as PySyft, Duet and PyGrid~\cite{Ryffel2018,PySyft,Duet}. Our focus is to extend the Hyperledger Aries functionalities, the~libraries designed for ML communication, and~distribute this framework as open-source to the academic and industrial community on the PyDentity project~\cite{PyDentity}. We hope that our work can motivate more people to work on the same subject. Additionally, our scope is to incorporate and evaluate further PPML techniques to create a fully trusted and secure environment for ML~computations.
\vspace{6pt}

\authorcontributions{All authors contributed to the manuscript's conceptualisation and methodology; P.P., W.A. and N.P. contributed in writing; P.P. performed the security evaluation of the proof-of-concept; W.A. developed the proof-of-concept's used libraries; A.J.H. performed the data preparation; W.J.B. reviewed and edited the manuscript. All authors have read and agreed to the published version of the~manuscript.}

\funding{The research leading to these results has been partially supported by the European Commission under the Horizon 2020 Program, through funding of the SANCUS project (G.A. n 952672).}



\dataavailability{Publicly available datasets were analyzed in this study. This data can be found here: \url{https://www.kaggle.com/osmi/mental-health-in-tech-survey} (accessed on 1 March~2021).} 

\conflictsofinterest{The authors declare no conflict of~interest.}

\appendixtitles{yes}
\appendixstart
\appendix
\section{Security~Testing}
\label{appsecuritytesting}

Traffic exchanged across DIDComm channel is always encrypted, as~in Figure~\ref{figencrypted}. Only the internal traffic during the connection establishment in a single network, such as information exchange between the hospital's agent and controller is not encrypted, as~shown in Figure~\ref{figproofs}. This is not considered an issue since we separated the agent and controller entities to simulate a real-world scenario in which those entities are individual~machines.

\end{paracol}
\nointerlineskip
\appendix
\begin{figure}[H]
\widefigure
\includegraphics[width=1\linewidth]{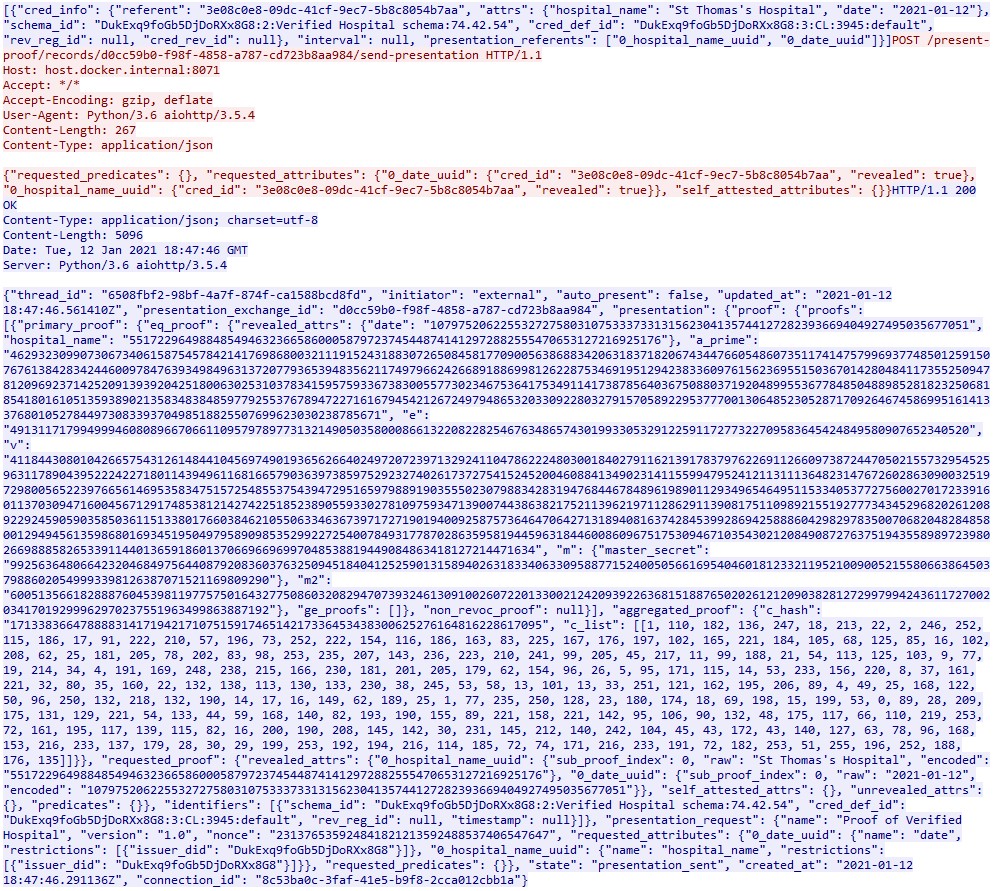}
\caption{Communication between agent and controller is not encrypted during the connection~establishment.}
\label{figproofs}
\end{figure}
\begin{paracol}{2}
\switchcolumn

\appendix
\begin{figure}[H]
\centering
\includegraphics[width=1\linewidth]{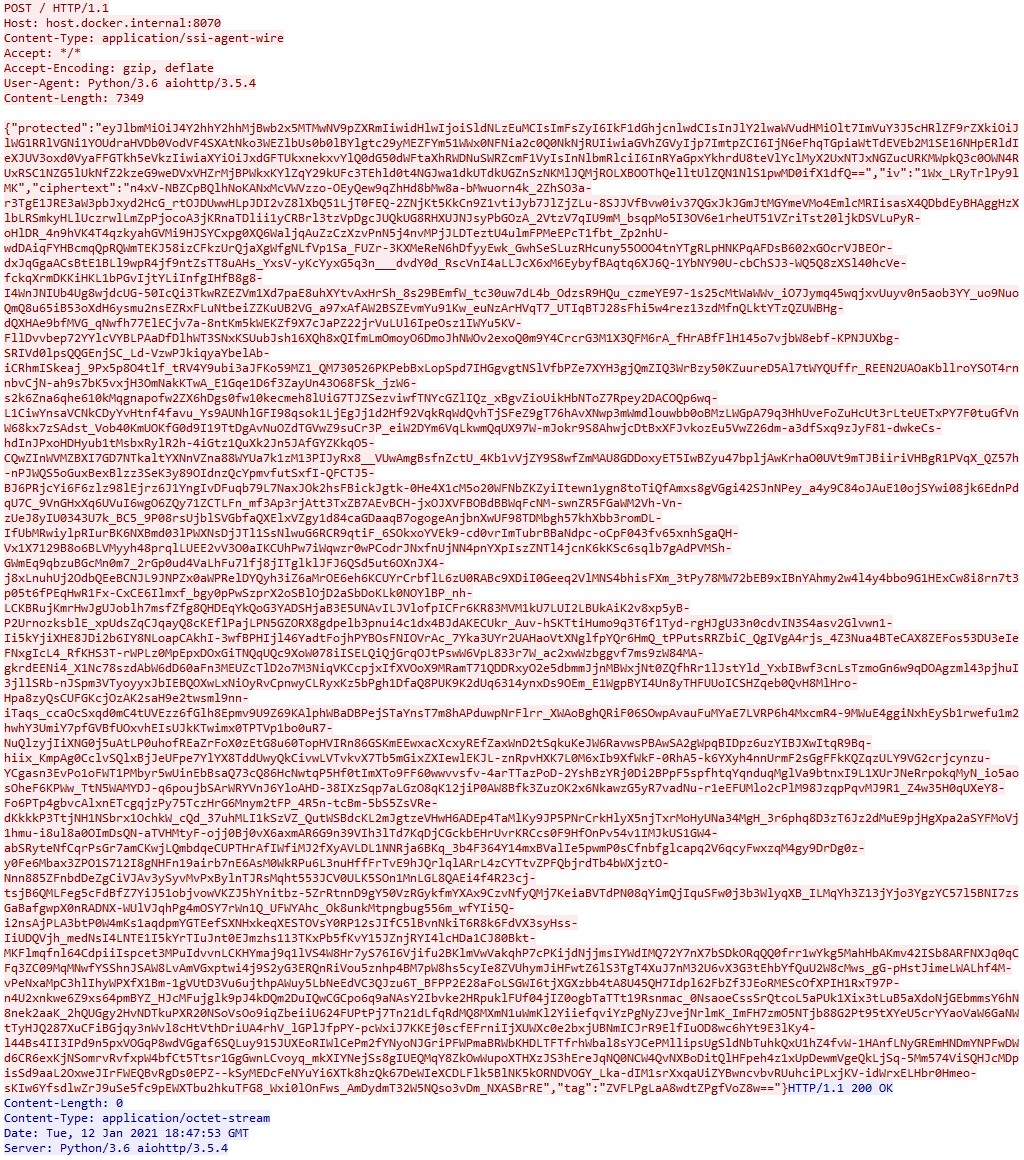}
\caption{Traffic through the DIDComm protocol is~encrypted.}
\label{figencrypted}
\end{figure}

\end{paracol}
\reftitle{References}

\end{document}